\documentclass[secnumarabic,amssymb, nobibnotes, pra, superscriptaddress, preprint]{revtex4-1}

\usepackage{lineno}
\usepackage{comment}

\usepackage{graphicx}
\usepackage{dcolumn}
\usepackage{bm}
\usepackage{amssymb}
\usepackage{amsmath}
\usepackage{longtable}
\usepackage{url}

\usepackage[english]{babel}

\usepackage{amsfonts}

\usepackage[left=3cm,right=2cm,top=2cm,bottom=2cm]{geometry}
\usepackage{graphicx}

\addto\captionsenglish{}

\begin{document}

\title{Single, double, and triple Auger processes in C$^{1+}$ for attosecond spectroscopy}

\author{Valdas Jonauskas} 
\email[]{Valdas.Jonauskas@tfai.vu.lt}
\author{\v{S}ar\={u}nas~Masys}

\affiliation{Institute of Theoretical Physics and Astronomy, Vilnius University, Saul\.{e}tekio av. 3, LT-10257 Vilnius, Lithuania}

\date{\today}

\begin{abstract}

We predict that attosecond spectroscopy could be used to observe time delay for electron emission {in the C$^{1+}$ ion} from a few step processes {that include electron impact excitation}. The results reveal  that the photon energy corresponding to the excitation to the $1s 2s^{2} 2p^{2} \, ^{2}D_{1.5}$ level from the ground one has to be used to study the emission of the electrons produced with the largest probability from a few step processes. 
Double and triple Auger processes are investigated in the C$^{1+}$ ion as a sequence of single Auger transition with subsequent ionization by the Auger electron.  Single- and two-step processes describe the double Auger transition while the triple Auger transition arises from two- and three-step processes. Fairly good agreement with experimental values for cross sections demonstrates that inner shell excitation by photon leads to subsequential emission of the electrons in the double and triple Auger processes. 
 
\end{abstract}

\maketitle

\section{INTRODUCTION} 

Fast multiple electron processes keep fascinating researchers for nearly 100 years. Understanding of such processes is essential for further development of theory and paving the road for technological advances \cite{2009rmp_81_163_krausz}. Ultrafast attosecond spectroscopy has been successfully applied to study motion of electrons in atoms \cite{2002n_419_803_drescher, 2014np_10_207_mansson, 2017np_13_280_ossiander}, molecules \cite{2014s_346_336_calegari, 2015s_350_6262_kraus}, and solids \cite{2014s_346_1348_schultze, 2017s_357_1274_siek}.  Interaction of photon with atom followed by multiple electron emission is attracting considerable interest due to  correlation effects which are of fundamental importance. Inner shell excitation or ionization can lead to emission of many electrons producing atoms in the highly ionized stages \cite{2010n_466_56_young, 2012np_6_858_rudek, 2018nc_9_719_hutten}. The main decay mechanism of the system with inner shell vacancy is Auger cascade when in every step of the cascade one electron is ejected from the ion. The cascade continues while energy of the atomic system is above the ionization threshold. The single Auger transition has the largest probability compared to the Auger transitions with the simultaneous emission of two \cite{1965prl_14_390_carlson} or three \cite{2008prl_100_233202dephilippo} electrons. However, simultaneous emission of multiple electrons is less understood and, therefore, more interesting as the description of the process has to deal with many body Coulomb problem. 

Time delay for photoionization from different atomic orbitals was experimentally observed using ultrafast attosecond spectroscopy in the Ne atom \cite{2010s_328_1658_schultze}. This effect was explained by the correlated motion of all electrons which occur during photoionization process. The time delay was also observed for the photoemission of electron in the He atom when ion is left in the shake-up state compared to the ground one \cite{2017np_13_280_ossiander}. Sudden perturbation of the atomic potential due to remove of one electron from the system leads to the rearrangement of the atomic wavefunction resulting in the excitation of the bound electron. On the other hand, the excited final state observed after photoionization can be explained by the interaction of the photoelectron with the bound electron on its way out from the ion. This \textit{additional process} would also lead to the time delay of the photoelectron leaving the ion in the excited state.  Unfortunately, it is impossible to distinguish which of the interpretations are realized in nature for this case. 

The attosecond spectroscopy can be applied to investigate this fundamental question related to \textit{additional processes} using  single, double, and triple Auger  transitions arising after the $1s \rightarrow 2p$ excitation in C$^{1+}$. These transitions were observed experimentally  by M\"{u}ller \textsl{et al.} \cite{2015prl_114_013002_muller, 2018pra_97_013409_muller}. It should be noted that double and triple Auger transitions can be of the largest importance for the $K$ shell decay in elements potentially applicable for targeted cancer therapies \cite{2000s_287_1658_boudaiffa, 2008prl_100_1_zheng, 2015nm_14_861_sanche}. 

Recently description of single, double, and triple Auger decay using the knock-out model was proposed  \cite{2016pra_93_060501_zhou}. They analyzed additional single and double ionization by electron produced in the single Auger process. Their work is based on the knock-out and shake-off mechanisms used before to describe the double Auger transitions \cite{1992pra_7_4576_amusia, 2013pra_87_033419_zeng}. The probabilities for the sequential ionization were taken to be equal to the collision strengths of the inelastic scattering for the knock-out modeling. The obtained ratios for triple to single Auger rates are above experimental values by about 40 \% but still within the experimental uncertainty of 50 \% \cite{2015prl_114_013002_muller}.  Only sequential single- and two-step direct ionization by Auger electron impact has been studied in this work to describe double and triple Auger transitions.

\begin{figure}
 \includegraphics[scale=0.7]{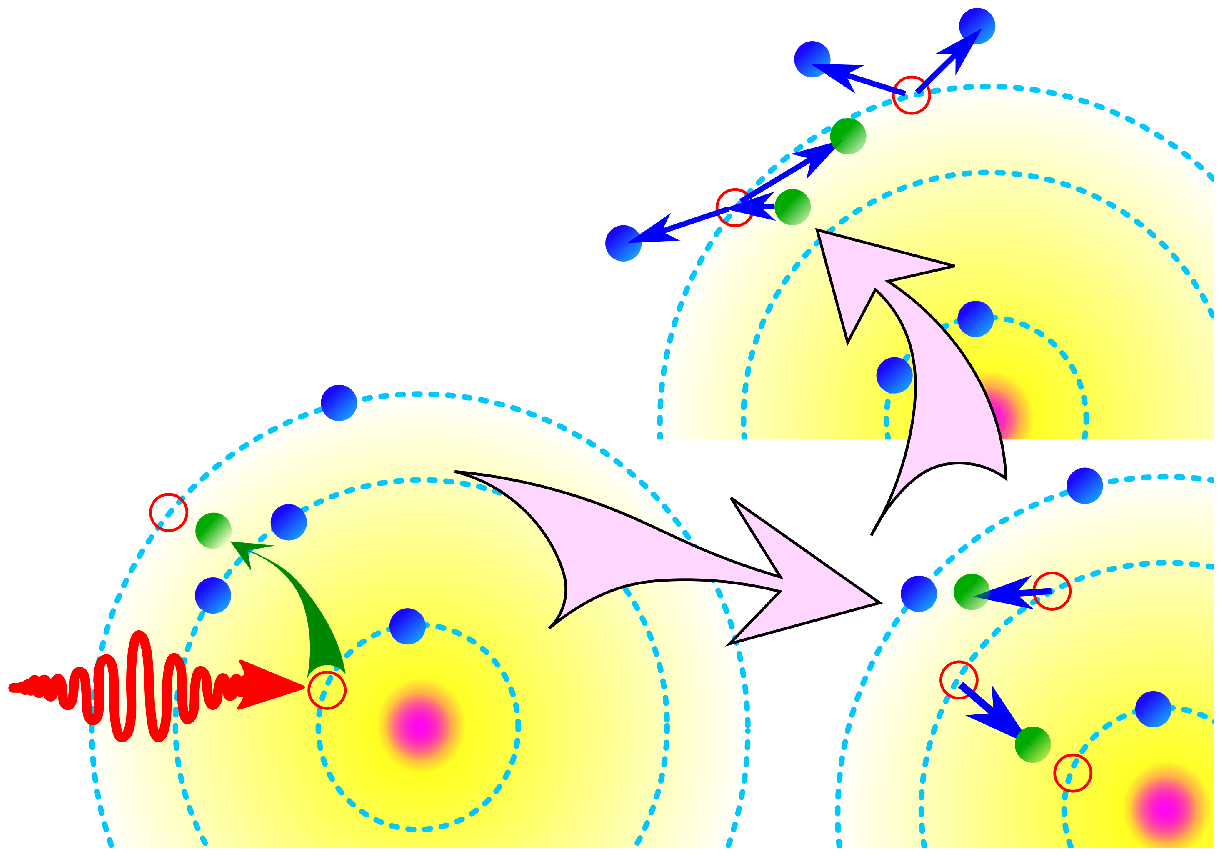}%
 \caption{\label{f0} (Color online)  Schematic illustration of the triple Auger transition after photoexcitation. See the text for the details.} 
\end{figure}

In this work, we demonstrate that time delays of emitted electrons {due to sequential collisions leading to} the double and triple Auger transitions for the C$^{1+}$ ion can be observed using the attosecond spectroscopy. { Every additional step in the sequential collision of the electrons results on average in the time delays for the emission process. } We suggest possible photon energies to study the effect in  C$^{1+}$. Emission of two and three Auger electrons is investigated as a consequence of interaction of single Auger electron with the bound electrons (Fig. \ref{f0}). This interaction can lead to electron-impact excitation or ionization. The interaction of Auger electron which results in ejection of one or two additional electrons from the system corresponds to the double or triple Auger process, respectively. Thus, the double Auger process is described by the electron-impact ionization or excitation with the subsequent ionization while the triple Auger process is presented by direct double ionization (DDI). Recently, the electron-impact DDI processes  were successfully investigated using a few step approach \cite{2014pra_89_052714_jonauskas, 2016pra_93_022711_konceviciute, 2018pra_97_012705_konceviciute}. It should be noted that the triple ionization by electron impact in the Se$^{2+}$ ion was explained as a sequence of DDI with subsequent autoionization \cite{2018pra_97_012705_konceviciute}. Interestingly, three-step DDI processes play a crucial role in the triple ionization of Se$^{2+}$. This result has further strengthened our hypothesis that the \textit{additional processes} that we have introduced {in addition to sequential ionization} are important in the multiple ionization. {Influence of these processes to the double and triple Auger transitions has not been analyzed before. }

{The rest of the paper is organized as follows. Section II presents a brief outline of the theoretical approach. In section III, the obtained results are discussed.
Finally, we end with the conclusions from the present investigation. }

\section{THEORETICAL APPROACH}

Cross sections of single, double, and triple ionization of the ground state C$^{1+}$ ion by photon with energy $h\nu$ are studied as a sequence of a few processes.  The ionization cross section $\sigma_{i f}(h\nu)$ from the initial level $i$ to the final level $f$ is expressed by the photoexcitation cross section $\sigma_{ij}^{ph} (h\nu)$  from the level $i$ to level $j$ of the C$^{1+}$ ion with subsequent autoionization to the level $k$ of C$^{2+}$ which can be followed by additional processes. This can be expressed by equation:
\begin{linenomath*}
\begin{equation}
\sigma_{i f}(h\nu) = \sum_{jk} \sigma_{ij}^{ph} (h\nu) \frac{A_{jk}^{\mathrm a}}{\sum_{m} A_{jm}^{\mathrm a}+\sum_{n} A_{jn}^{\mathrm r}} P_{kf}(\varepsilon).
\label{e1}
\end{equation} 
\end{linenomath*}
Here summation is performed over the intermediate levels $j$ and $k$, $A^{a}$ and $A^{r}$ are the Auger and radiative transition probabilities, respectively. The summation in the denominator determines all decays from the level $j$ through the Auger and radiative transitions. 
The Auger electron of the energy $\varepsilon$ on its way out from the system can eject one or two additional electrons with the corresponding probability $P_{kf}(\varepsilon)$ leading to the transition to the C$^{3+}$ or C$^{4+}$ ionization stages, respectively. 

For single ionization by photon, the probability $P_{kf}(\varepsilon)$ is expressed by Kronecker delta: 
\begin{linenomath*}
\begin{equation}
P_{kf}(\varepsilon) = \delta_{kf}.
\label{e2}
\end{equation} 
\end{linenomath*}
This means that the Auger electron leaves system of the bound electrons without further interaction. 
For double ionization, the probability includes electron-impact collisional ionization (CI) and excitation with subsequent ionization (EI):
\begin{linenomath*}
\begin{equation}
P_{kf}(\varepsilon) = P_{kf}^{\mathrm{ CI}}(\varepsilon) + P_{kf}^\mathrm{ EI}(\varepsilon).
\label{e3}
\end{equation} 
\end{linenomath*}
The probability for CI is obtained from equation: 
\begin{linenomath*}
\begin{equation}
P_{kf}^\mathrm{ CI}(\varepsilon) = \frac{\sigma_{kf}^\mathrm{ CI}(\varepsilon)}{4\pi\bar{R}_{nl}^2},
\label{e4}
\end{equation} 
\end{linenomath*}
where \(\bar{R}_{nl}\) is the mean distance of the electrons in the \(nl\) shell from the nucleus and the electron is ejected from the \(nl\) shell. The probability for EI process is given by
\begin{linenomath*}
\begin{equation}
P_{kf}^\mathrm{ EI}(\varepsilon) = \sum_{p} \frac{\sigma_{kp}^\mathrm{ CE}(\varepsilon)}{4\pi\bar{R}_{nl}^2} \frac{\sigma_{pf}^\mathrm{ CI}(\varepsilon_1)}{4\pi\bar{R}_{n'l'}^2},
\label{e5}
\end{equation} 
\end{linenomath*}
where the level $p$ belongs to the C$^{2+}$ ion, $\varepsilon_1 = \varepsilon - \bigtriangleup E_{kp}$, $\bigtriangleup E_{kp}$ is the transition energy. 

For the triple ionization, the probability takes into account ionization-ionization (II), excitation-ionization-ionization (EII), and ionization-excitation-ionization (IEI) processes
\begin{linenomath*}
\begin{equation}
P_{kf}(\varepsilon) = P_{kf}^\mathrm{ II}(\varepsilon) + P_{kf}^\mathrm{EII}(\varepsilon)  + P_{kf}^\mathrm{ IEI}(\varepsilon).
\label{e6}
\end{equation} 
\end{linenomath*}
These processes define the probabilities for DDI \cite{2014pra_89_052714_jonauskas, 2016pra_93_022711_konceviciute, 2018pra_97_012705_konceviciute}. 
The probability of the II process is expressed by equation 
\begin{linenomath*}
\begin{equation}
P_{kf}^\mathrm{ II}(\varepsilon) = \sum_{r} \frac{\sigma_{kr}^\mathrm{ CI}(\varepsilon)}{4\pi\bar{R}_{nl}^2} \frac{\sigma_{rf}^\mathrm{ CI}(\varepsilon_1)}{4\pi\bar{R}_{n'l'}^2},
\label{e7}
\end{equation} 
\end{linenomath*}
where the level $r$ corresponds to the C$^{3+}$ ion, $\varepsilon_1$ is the energy of the ejected or scattered electron after ionization from the level $k$.
The probability of the EII process is given by product of three probabilities
\begin{linenomath*}
\begin{equation}
P_{kf}^\mathrm{ EII}(\varepsilon) = \sum_{pr} \frac{\sigma_{kp}^\mathrm{ CE}(\varepsilon)}{4\pi\bar{R}_{nl}^2} \frac{\sigma_{pr}^\mathrm{ CI}(\varepsilon_1)}{4\pi\bar{R}_{n'l'}^2} \frac{\sigma_{rf}^\mathrm{ CI}(\varepsilon_2)}{4\pi\bar{R}_{n''l''}^2}.
\label{e8}
\end{equation} 
\end{linenomath*}
Here $\varepsilon_1 = \varepsilon - \bigtriangleup E_{kp}$, $\varepsilon_2$ is the energy of the ejected or scattered electron after the ionization from the level $p$.
In the same way, the probability of the IEI process can be expressed by equation
\begin{linenomath*}
\begin{equation}
P_{kf}^\mathrm{ IEI}(\varepsilon) = \sum_{tr} \frac{\sigma_{kt}^\mathrm{ CI}(\varepsilon)}{4\pi\bar{R}_{nl}^2} \frac{\sigma_{tr}^\mathrm{ CE}(\varepsilon_1)}{4\pi\bar{R}_{n'l'}^2} \frac{\sigma_{rf}^\mathrm{ CI}(\varepsilon_2)}{4\pi\bar{R}_{n''l''}^2},
\label{e9}
\end{equation} 
\end{linenomath*}
where the level $t$ belongs to the C$^{3+}$ ion, $\varepsilon_1$ is the energy of the ejected or scattered electron after the first electron-impact ionization, $\varepsilon_2 = \varepsilon_1 - \bigtriangleup E_{tr}$. 

Energy levels, radiative and Auger  transition probabilities  as well as electron-impact excitation and CI cross sections have been calculated using the Flexible Atomic Code \cite{2008cjp_86_675_Gu}. This code implements the Dirac-Fock-Slater approach.  Electron-impact excitation and CI processes are investigated using the distorted-wave (DW) approximation.

\section{RESULTS}

Photoexcitation from the ground level C$^{1+}$ $1s^{2}2s^{2}2p\, ^{2}P_{0.5}$ leads to the formation of $1s2s^{2}2p^{2}\, ^{2}P_{0.5, 1.5}$ and $^{2}D_{1.5}$ levels. The excitation to the $1s2s^{2}2p^{2}$ $^{2}D_{2.5}$ level is forbidden due to restriction rule $\bigtriangleup J = 0, 1$ for the electric dipole transitions. 
Theoretical single ionization threshold for the C$^{2+}$ ion is 45.69 eV. This is slightly below the value of 47.89 eV provided by the NIST  \cite{NIST_ASD}. The theoretical double ionization threshold is 109.88 eV while the NIST recommended value amounts to 112.38 eV. 

\begin{figure}
 \includegraphics[scale=0.7]{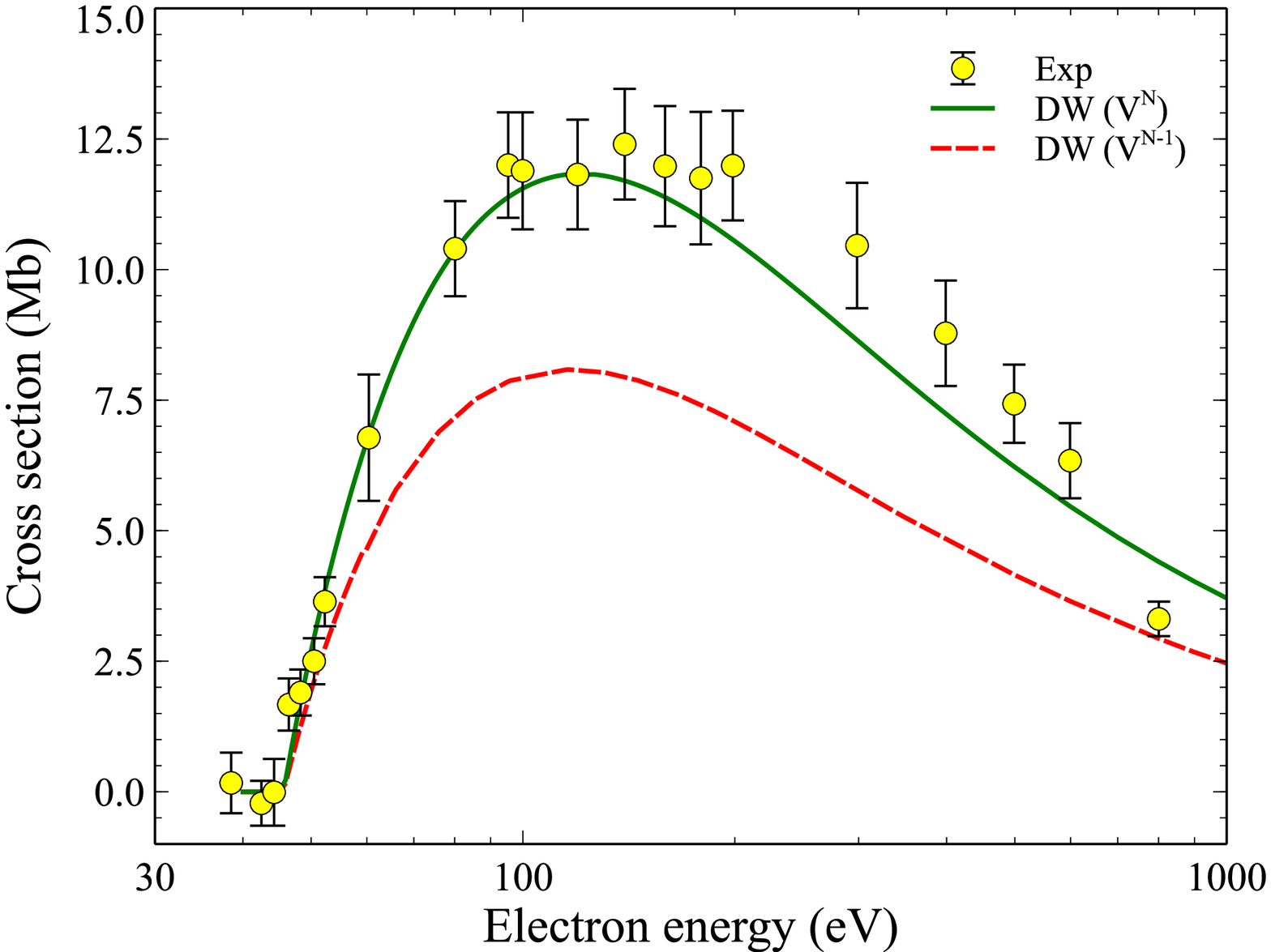}%
 \caption{\label{f1} (Color online)  Electron-impact single ionization cross sections for the ground level of the C$^{2+}$ ion. Data are obtained in the potential of ionizing (green solid) and ionized (red dashed) ion. Experiment: Exp \cite{2008apjss_175_543_fogle}. } 
\end{figure}

\begin{figure}
 \includegraphics[scale=0.7]{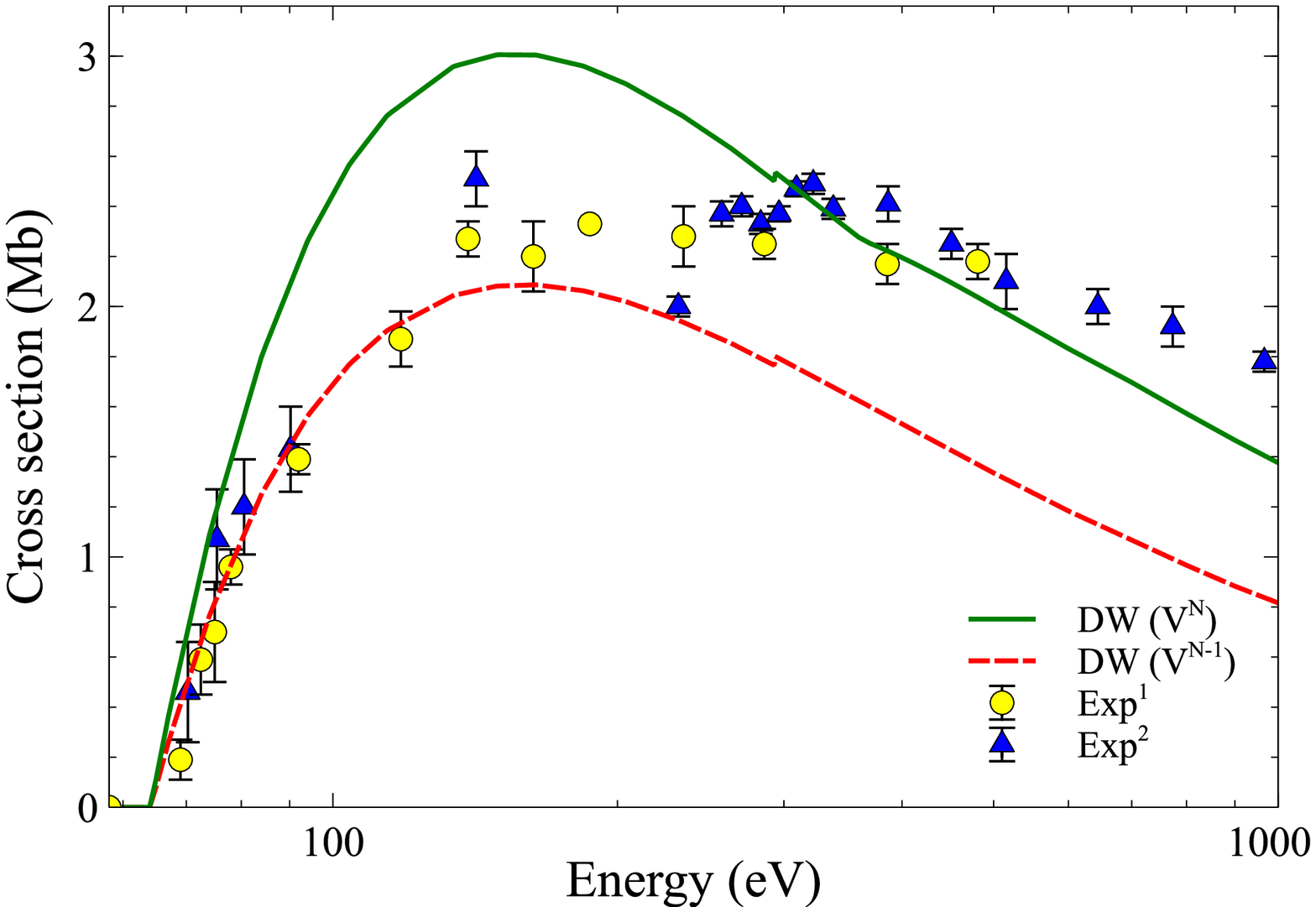}%
 \caption{\label{f2} (Color online)   Electron-impact single-ionization cross sections for the ground level of the C$^{3+}$ ion. Data are obtained in the potential of ionizing (green solid) and ionized (red dashed) ion. Experiment: Exp$^{1}$ \cite{1978pra_18_1911_crandall}, Exp$^{2}$ \cite{1979jpb_12_l249_crandall}. } 
\end{figure}

Single ionizations by electron-impact for the C$^{2+}$ and C$^{3+}$ ions have to be investigated since the study of  additional  emission of electrons after single Auger transition includes the ionization processes [Eqs. (\ref{e4}), (\ref{e5}), (\ref{e7}), (\ref{e8}), and (\ref{e9})]. 
Comparison with  experimental cross sections for the ionization from the ground levels of the C$^{2+}$ and C$^{3+}$ ions are presented in Figs. \ref{f1} and \ref{f2}. For the C$^{2+}$ ion (Fig. \ref{f1}),  better agreement with measurements \cite{2008apjss_175_543_fogle} is obtained when the ionization process is studied in the potential of ionizing ion ($V^{N}$). Thus, cross sections of the ionization by the initial Auger electron for the double Auger process have to be studied in the corresponding potential. However, the different situation is observed for the single ionization process in the C$^{3+}$ ion (Fig. \ref{f2}). At the lower energies of the incident electron, better agreement with the experiment \cite{1978pra_18_1911_crandall, 1979jpb_12_l249_crandall} is achieved when cross sections are investigated in the potential of the ionized ion ($V^{N-1}$). However, calculations in the potential of the ionizing ion ($V^{N}$) provide  better resemblance to the measurements for the higher energies. 

Energies of Auger electrons produced by the single Auger process vary from 239.73 to 266.83 eV. Since these Auger electrons participate in the subsequent ionization, cross sections of the ionization for the C$^{2+}$ ion have to be considered for the corresponding energies (Fig. \ref{f1}).  

Two limiting cases of energy distribution for electrons after the first ionization process are considered in our approach for the ionization from the C$^{3+}$ ion. The energy range from 194.04 to 221.14 eV is taken into account when the excess energy is taken by one of the ejected or scattered electrons after the first ionization process from the ground level of the C$^{2+}$ ion. This energy range corresponds to the Auger electron energy diminished by the single ionization threshold for C$^{2+}$. However, the energies of the electrons are lower if the ionization occurs to the excited levels of the C$^{3+}$ ion. For the situation when the electrons share the excess energy, the energy range from 97.02 to 110.57 eV has to be investigated for transitions to the ground level of C$^{3+}$. 
Again, these energies are lower for the ionization from and/or to the excited levels of C$^{2+}$ and C$^{3+}$.  
Our study of the double and triple Auger processes shows that better agreement with experiment is found when one of the electrons takes all the excess energy. Thus, the asymmetric distribution for electron energies is observed. Therefore, the energy range from 194.04 to 221.14 eV has to be considered. In this energy range,  better agreement with experiment for the CI cross sections is obtained in the potential of the ionized ion (V$^{N-1}$). 

\begin{figure}
 \includegraphics[scale=0.7]{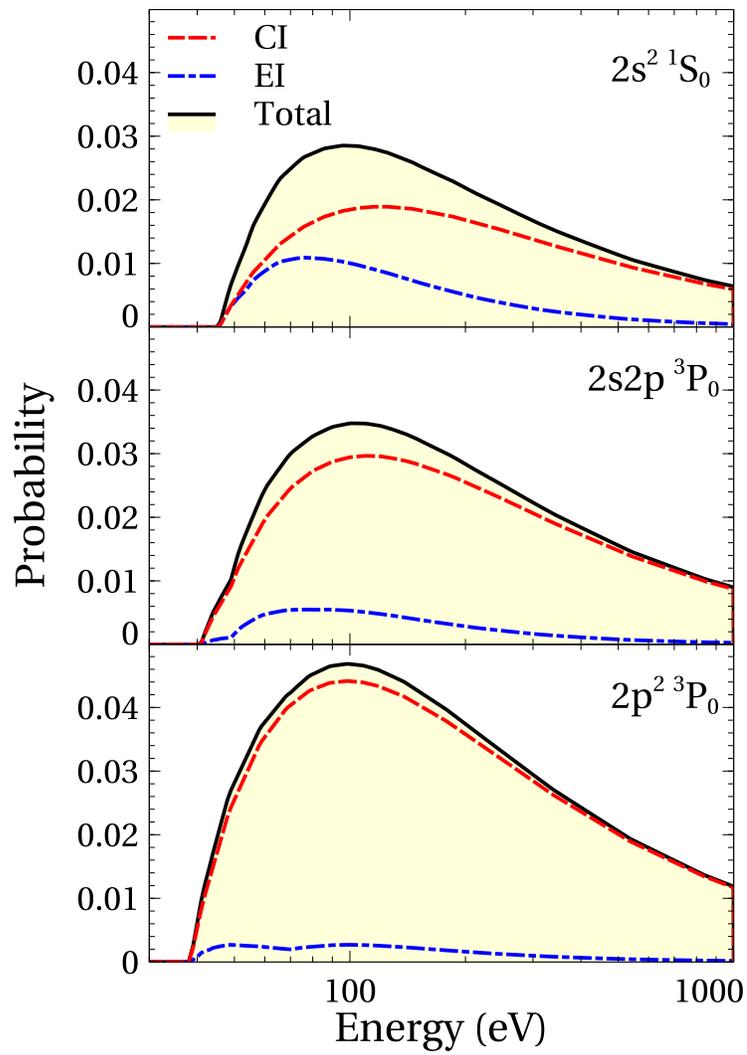}%
 \caption{\label{f3} (Color online)  Probabilities of the single- and two-step processes for double Auger transitions corresponding to additional ionization by the Auger electron from the lowest levels of the  C$^{2+}$  $2s^{2}$, $2s2p$, and $2p^{2}$ configurations. The total probabilities are presented by the black solid line. } 
\end{figure}

Double Auger process consists of single Auger transition with subsequent electron-impact ionization [Eq. (\ref{e4})] and excitation with ionization [Eq. (\ref{e5})]. Probabilities of the subsequent processes for the lowest levels of the $2s^{2}$, $2s2p$, and $2p^{2}$ configurations are presented in Fig. \ref{f3}. The most remarkable result to emerge from the data is that the two-step EI process provides quite large contribution. The contribution of the EI process consists of 35\% at the peak of the total probability for the $2s^{2}$ $^{1}S_{0}$ level. Its influence is even larger at the lower energies. The contribution of the process drastically decreases at the higher energies of the incident electron. The two-step process also plays less significant role for other configurations (Fig. \ref{f3}). It should be noted that the probability to form C$^{3+}$ is the largest for the C$^{2+}$ $2p^{2}$ configuration. It is higher by about 60 \% compared to the $2s^{2}$ configuration and 40 \% than the $2s2p$ configuration at the peak.  

\begin{figure}
 \includegraphics[scale=0.7]{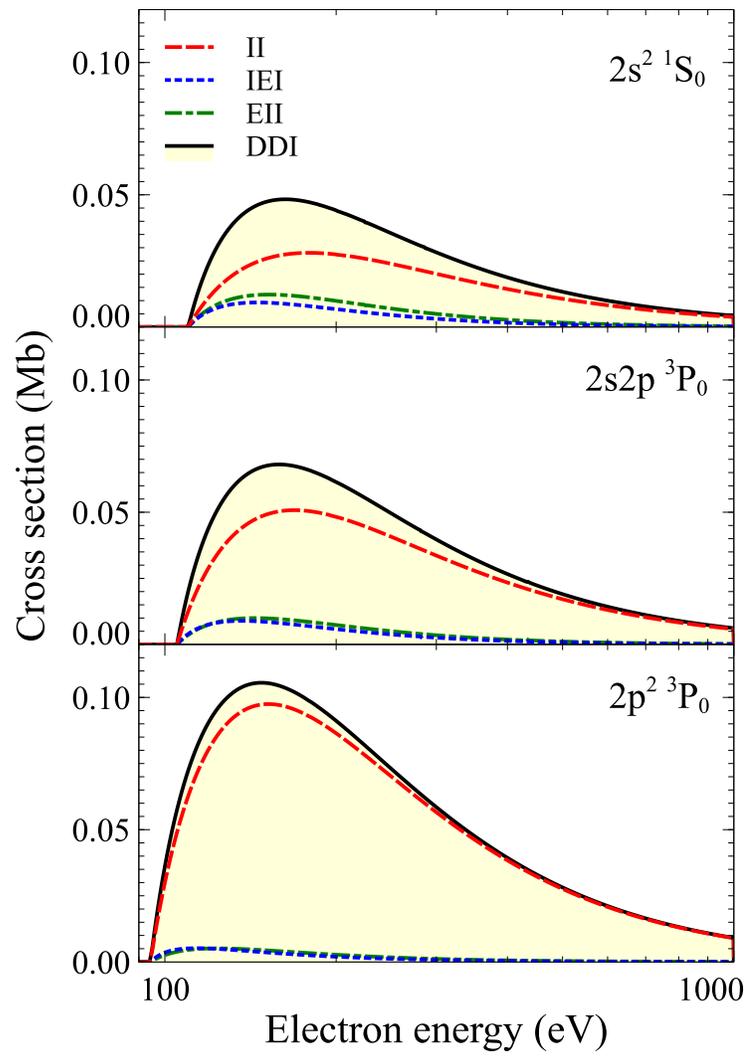}%
 \caption{\label{f4} (Color online)  Cross sections of the DDI process corresponding to additional ionization by Auger electron from the lowest levels of the  C$^{2+}$ $2s^{2}$, $2s2p$, and $2p^{2}$ configurations. The total cross sections are presented by the black solid line. } 
\end{figure}

\begin{figure}
 \includegraphics[scale=0.7]{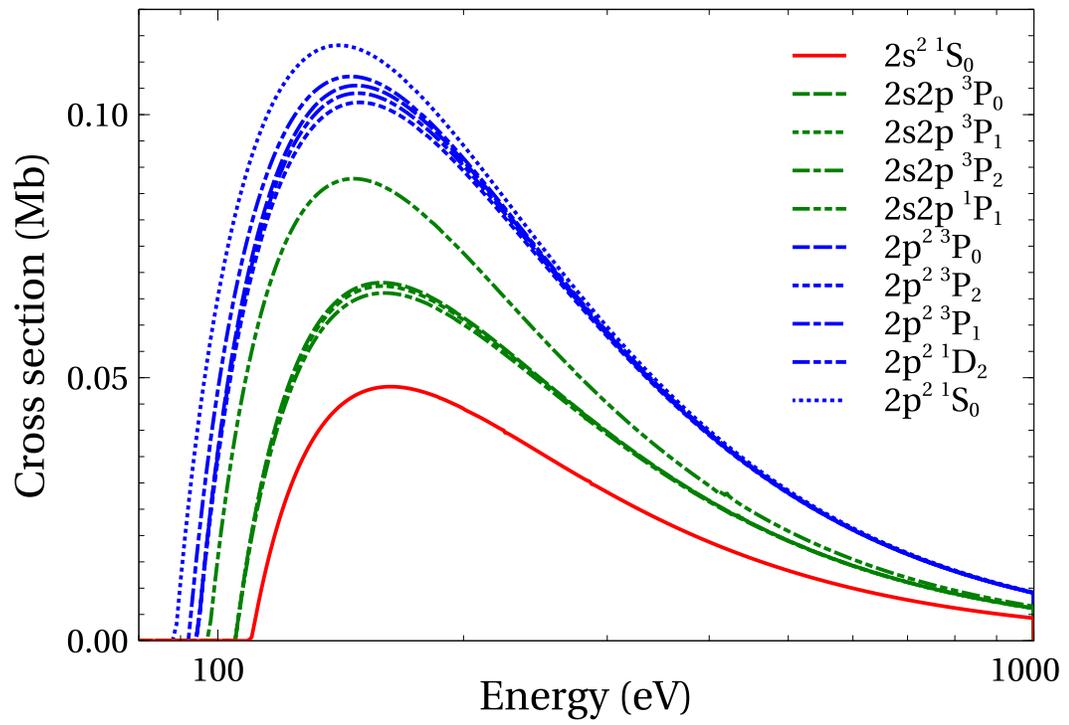}%
 \caption{\label{f5} (Color online)  Cross sections of the DDI process for all levels of the  C$^{2+}$ $2s^{2}$, $2s2p$, and $2p^{2}$ configurations. } 
\end{figure}

The DDI cross sections for the lowest levels of the $2s^{2}$, $2s2p$, and $2p^{2}$ configurations are shown in Fig. \ref{f4}. The data correspond to the case when one of the electrons takes all the excess energy after the first ionization process. Contribution from the two- and three-step processes is also presented. The obtained data demonstrate that II provides the largest input to the DDI process. The relative contribution of the three-step processes is the largest for the $2s^{2}$ $^{1}S_{0}$ level. On the other hand, the II process consists of 92 \% of the total DDI cross sections at the peak for the $2p^{2}$ $^{3}P_{0}$ level. At the higher energies, influence of the three-step processes is negligible. 
{The DDI cross sections from all levels of the $2s^{2}$, $2s2p$, and $2p^{2}$ configurations are presented in Fig. \ref{f5}. }
The calculations for all levels of the $2s^{2}$, $2s2p$, and $2p^{2}$ configurations show that the DDI cross sections are about two times higher for the $2p^{2}$ configuration compared to the $2s^{2}$ one. This can be explained by the fact that the CI cross sections are higher for the electron-impact ionization from the $2p$ shell compared to the $2s$ one. 

\begin{figure}
 \includegraphics[scale=0.7]{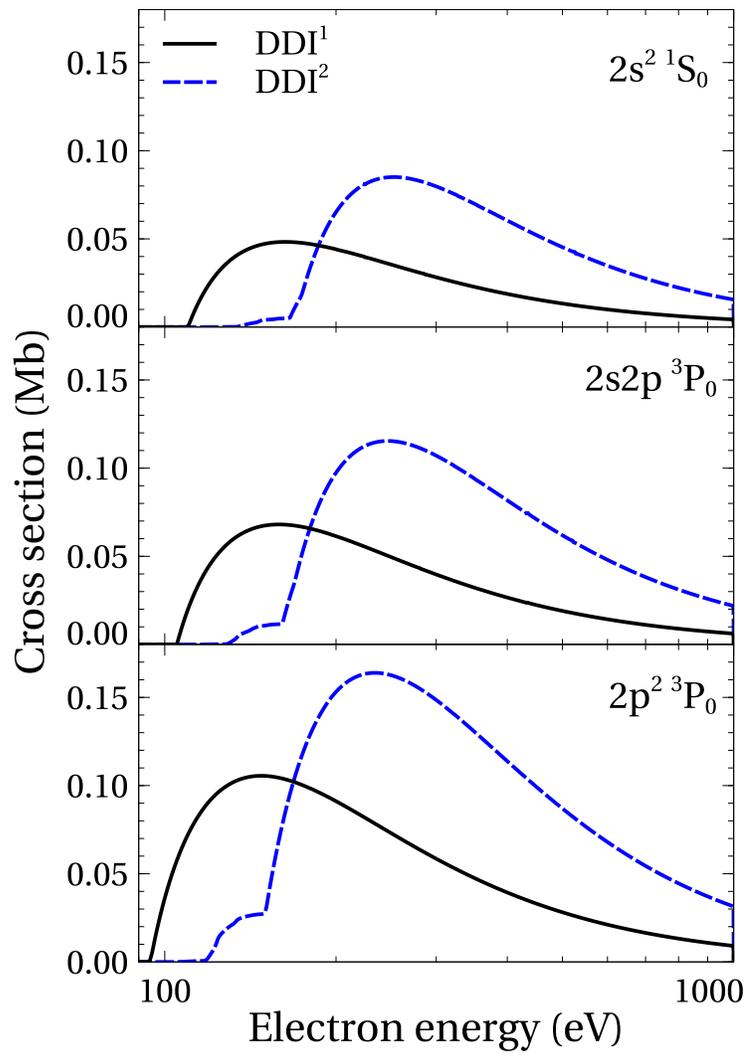}%
 \caption{\label{f6} (Color online)  Cross sections of the DDI process for the lowest levels of the C$^{2+}$ $2s^{2}$, $2s2p$, and $2p^{2}$ configurations. DDI$^{1}$ - one of the electrons takes all the excess energy after the first ionization process, DDI$^{2}$ - electrons share the excess energy. } 
\end{figure}

{The DDI cross sections when the ejected and scattered electrons share the excess energy after the first electron-impact ionization process are compared to the situation when one of the electrons takes all the excess energy in Fig. \ref{f6}. The first scenario dominates at the higher energies of the incident electron while the second case leads to the higher cross sections at the lower energies. }

\begin{figure}
 \includegraphics[scale=0.7]{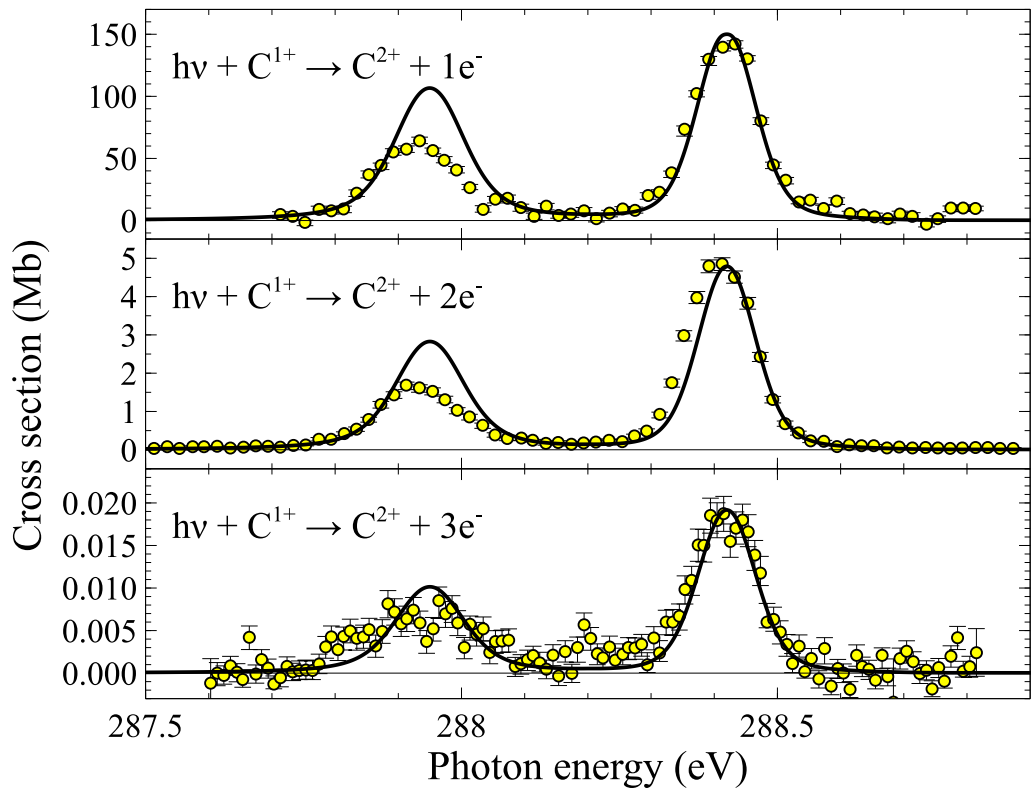}%
 \caption{\label{f7} (Color online)  Single, double, and triple ionization from the ground state of C$^{1+}$ by photon. Solid line (black): theoretical values, empty circles: experiment \cite{2015prl_114_013002_muller}. Theoretical values were convoluted with a Gaussian distribution function of 92 meV full width at half maximum.  } 
\end{figure}

Cross sections for single, double, and triple ionization by photon from the ground state of the C$^{1+}$ ion are compared with experimental values in Fig. \ref{f7}. The value of the main peak is normed to the experimental one for the cross sections of the single Auger process. In addition, the theoretical values are shifted to the measured peak positions.  The main peak corresponds to excitation from the C$^{1+}$ $1s^{2}2s^{2}2p$ $^{2}P_{0.5}$ level to the $1s2s^{2}2p^{2}$ $^{2}P_{0.5, 1.5}$ levels. The lower peak is produced by excitation to the  $1s2s^{2}2p^{2}$ $^{2}D_{1.5}$ level. Fairly good agreement with measurements {for the relative intensities of the single, double, and triple Auger transitions } demonstrates that these processes can be described by a few step model. Therefore, there has to be a time delay for electron emission from the different branches of the processes {since every additional collision leads to the delay in the electron emission}. 

\begin{figure}
 \includegraphics[scale=0.7]{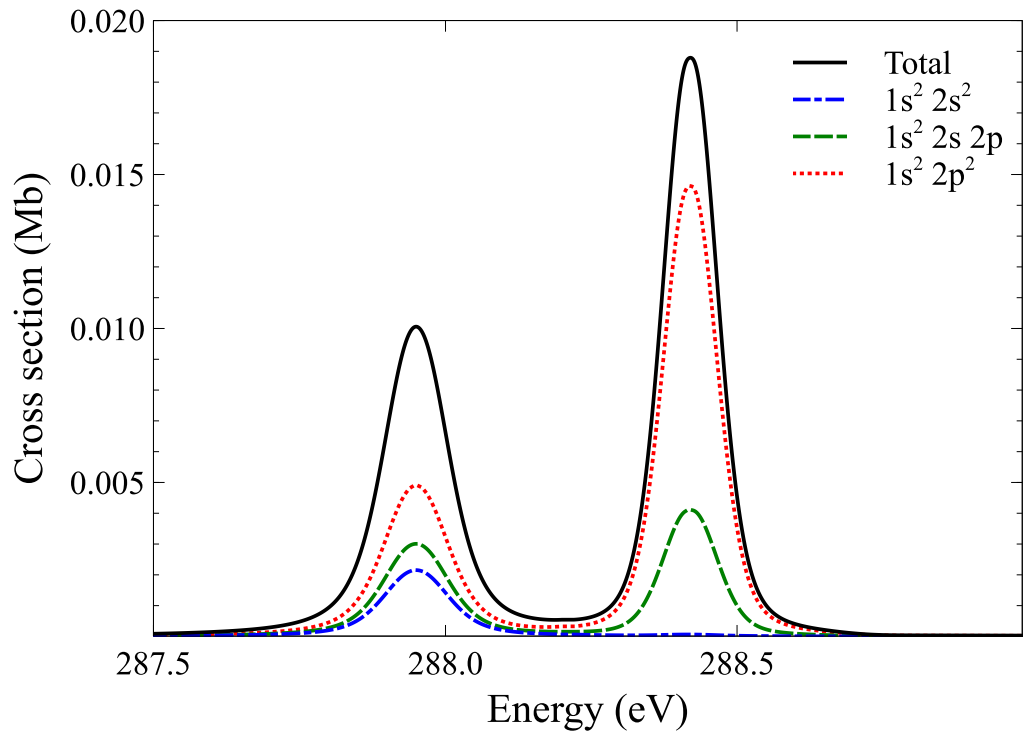}%
 \caption{\label{f8} (Color online)  Contribution of the  C$^{2+}$ $2s^{2}$, $2s2p$, and $2p^{2}$ configurations to the triple ionization from the ground state of C$^{1+}$ by photon. Cross sections were convoluted with a Gaussian distribution function of 92 meV full width at half maximum. } 
\end{figure}

Contribution to the triple ionization cross sections for the C$^{2+}$ $1s^{2}2s^{2}$, $1s^{2} 2s 2p$, and $1s^{2}2p^{2}$ configurations obtained by the single Auger decay from the  $1s2s^{2}2p^{2}$ $^{2}P_{0.5, 1.5}$ and $^{2}D_{1.5}$ levels is shown in Fig. \ref{f8}. It can be seen that influence of the $1s^{2}2p^{2}$ configuration dominates for both peaks. The cross sections produced by DDI from the $1s^{2}2s2p$ configuration are about 3.6 times smaller compared to the contribution from the $1s^{2}2p^{2}$ configuration for the main peak. On the other hand, input from the $1s^{2}2s^{2}$ configuration is not seen at this energy range. This can be explained by the fact that corresponding submatrix element $\langle 1s 2s^{2} 2p^{2} \, {}^{2}P \| H^{e} \| 1s^{2} 2s^{2} \, {}^{2}S \varepsilon p \, {}^{2}P \rangle$ of the two-electron Coulomb interaction Hamiltonian $H^{e}$ is zero. Contribution to the triple ionization from the $1s^{2}2s^{2}$ configuration is observed for the lower peak only. As it is demonstrated above, the influence of the three-step processes to the formation of the DDI cross sections is the largest for the $1s^{2}2s^{2}$ configuration. The same is true for the double Auger process but in this case the time delay for the two-step process compared to the single-step one could be observed.

\section{CONCLUSIONS} \label{conclusions}

In conclusion, the attosecond spectroscopy can be applied to observe  time delays in electron emission after the $1s \rightarrow 2p$ excitation in the C$^{1+}$ ion. The largest probability to observe the two-step process for the double Auger transition corresponds to the $1s^{2}2s^{2}2p \, ^{2}P_{0.5} \rightarrow 1s2s^{2}2p^{2} \, ^{2}D_{1.5}$ excitation. The obtained results reveal that this process would provide the signal by an order of magnitude higher compared to transition to the next ionization stage. The same energy range has to be used to observe the time delays from the three-step processes compared to the two-step one for the triple Auger transitions. Combined the time delay measurements for the electron emission with the ion spectroscopy would result in the separation of the single, double, and triple Auger processes. It would help to resolve the long standing problem about the role of a few step processes in the so called simultaneous emissions of multiple electrons. 





\newpage

\end{document}